# Wi-Fi Wardriving Studies Must Account for Important Statistical Issues

Edward J Oughton, Julius Kusuma, Thibault Peyronel, and Jon Crowcroft, *Fellow*, IEEE

**Abstract**— Knowledge of Wi-Fi networks helps to guide future engineering and spectrum policy decisions. However, due to its unlicensed nature, the deployment of Wi-Fi Access Points is undocumented meaning researchers are left making educated guesses as to the prevalence of these assets through remotely collected or passively sensed measurements. One commonly used method is referred to as 'wardriving' essentially where a vehicle is used to collect geospatial statistical data on wireless networks to inform mobile computing and networking security research. Surprisingly, there has been very little examination of the statistical issues with wardriving data, despite the vast number of analyses being published in the literature using this approach. In this paper, a sample of publicly collected wardriving data is compared to a predictive model for Wi-Fi Access Points. The results demonstrate several statistical issues which future wardriving studies must account for, including selection bias, sample representativeness and the modifiable areal unit problem.

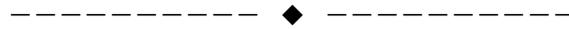

## 1. Introduction

The network of Wi-Fi Access Points (APs) that so many of us use daily is an essential part of the digital connectivity required by society. In many places Wi-Fi is the primary technology used to connect to the Internet. Over the last decade Internet-connected devices such as smartphones, tablets, home devices and wearable electronics have been widely adopted, helping to drive the usage of Wi-Fi technologies as a low-cost way to transport vast quantities of data [1]. Thanks to Wi-Fi using unlicensed spectrum the deployment of these assets is cheaper and easier than competing technologies such as cellular [2]. Indeed, Wi-Fi certified devices are designed to comply with spectrum regulations, including transmit power, spectrum utilization, channel allocations, and protocols for multiple access. These protocols are designed to enable multiple Wi-Fi APs and devices to be deployed nearby each other and manage interference [3].

Wireless communication systems are used for a very large number of applications, from smartphone and smart home devices [4], to monitoring infrastructure assets, supply chains, the environment and vehicle connectivity [5]–[12]. The use of wireless networks is still rapidly evolving, with new use cases ranging from holographic video to augmented or virtual reality (AR/VR) [13].

Understanding the spatial structure of self-deployed wireless networks is invaluable for design and management, including knowledge on how densely Wi-Fi APs are deployed. This insight is essential for network simulations which test new architectures or protocols, and ensuring that 'what looks good on paper' actually lives up to reality when deployed in real world networks [14]. For example, recent contributions focusing on Wi-Fi density include the cooperative control of Wireless Local Area Networks (WLANs) using IEEE 802.11ax, providing a capacity improvement of approximately 11% [15]. Moreover, information on Wi-Fi assets can help to inform bandwidth partitioning in ultra-dense wireless networks [16].

However, as Wi-Fi is generally setup and managed locally, these network assets are like other decentralized areas of the Internet in that they are undocumented and can often only be guessed-at through remote measurement [17]. The method of 'wardriving', 'warbiking', 'wardroning' or 'warwalking', has therefore emerged as a way for computer scientists and communications engineers to actively collect data on wireless networks by passively sensing deployed equipment. The different forms of mobility, either a car, bicycle, drone or merely on foot, each have their own nuanced impact on the results obtained [18]. The term wardriving is said to have emerged from the 1980s movie WarGames where 'wardialing' was used to dial telephone numbers in search of modems [19].

With the growing importance of Wi-Fi technology for serving the vast expansion of Internet access traffic in a cost-efficient way, it is essential to have a systematic understanding of the prevalence and use of Wi-Fi for both network and security researchers (and beyond).

However, our conjecture in this paper is that we should be much more concerned with how wardriving exercises are designed and data collected, as this has a significant impact on the scientific robustness of the results. Incorrect conclusions could be drawn if the statistical limitations of this method are not appropriately controlled for.

A review of Wi-Fi data is undertaken next, before the method is presented in Section 3 and results reported in Section 4. Various statistical issues are discussed in Section 5 before Section 6 provides conclusions.

- *E. Oughton is with the College of Science, George Mason University, Fairfax, VA, 22030. E-mail: eoughton [at] gmu [dot] edu*
- *J. Kusuma and T. Peyronel are with Facebook Connectivity Lab, Menlo Park, CA, 94025.*
- *J. Crowcroft is with the Computer Laboratory, University of Cambridge, Cambridge, UK, CB3 0FD.*

## 2.  Wi-Fi data sources and collection methods

Wi-Fi APs exhibit spatial statistical patterns, which make these technologies amenable to predictive modeling [20], such as predicting Wi-Fi hotspots [21]. There are four main use cases for this type of activity. Firstly, estimating coverage area scenarios by exploring received signal strength, application usage and traffic patterns [22], [23]. Secondly, assessing hotspots in range, enabling the exploration of scenarios for handovers, interference and bandwidth resource sharing [24], [25]. Thirdly, experimenting with Wi-Fi mesh network designs and backhaul traffic optimization, given such concepts rely on communication among hotspots [26]–[29]. Finally, developing Wi-Fi inventories to inform network management, location sensing [30] and detect unauthorized Access Points [31].

*Crowdsourced wardriving* has grown over the past decade with the emergence of smartphones making it easier and lower cost to undertake surveys, leading to a vast expansion in the number of studies using this approach. Underpinning this collection method is the availability of free to use smartphone applications. One of the most popular is the Wireless Geographic Logging Engine from WiGLE.net, which enables users to collect and download data for wireless networks detected using the app [32]. Collected data contains a range of parameters which can allow mapping of Wi-Fi APs and other types of wireless nodes [33], [34]. An advantage of WiGLE is open access to its API, allowing users to download spatially explicit crowdsourced Wi-Fi data with few usage limitations [35]. A drawback from this approach is that the number of daily API queries allowed is very limited.

From a mobile computing perspective, a key use of this approach is to map, quantify and analyze Wi-Fi Access Point density in specific areas [36]. Such research activities originally began with people collecting data on Wi-Fi network security to inform cyber security research [37]–[40], but many other uses for the technique have emerged. These include identifying vacant homes to inform urban planning policies [41], or fighting crime and terrorism [42]. In contrast, publicly generated Wi-Fi Access Point data can be obtained by undertaking either *targeted* or *opportunistic* wardriving using the WiGLE app. There are advantages to each approach.

*Targeted wardriving* can provide a near-comprehensive understanding of Wi-Fi Access Points for surveyed areas but can be incredibly time intensive and therefore it is hard to survey large areas. Indeed, without using drones targeted wardriving is usually confined to the street level, which means only approximately the bottom two floors of a building are sensed. The main benefit from this approach is the ability to methodically design a collection regime that controls factors which affect Wi-Fi sensing, such as the collection device and speed of travel. However, there are wardriving assessments which have used this targeted approach and then attempted to generalize the results from a smaller study area to a wider context [43]–[46]. Such an approach does not provide a suitably representative sample to allow such a generalization to be made, unless the sample size can be demonstrated to meet required statistical sample size rules which aim to minimize bias. Moreover, wardriving assessments often do not consider how the results of an application may statistically generalize to other contexts [42]. More robustly designed wardriving samples define a study area and then attempt to collect data across as much of the area as possible (although often falling short of providing quantified statistics on the percentage of the surveyed area) [47]–[49], [49], [50].

*Opportunistic wardriving*, in contrast, can benefit from frequent and inexpensive gathering of network data from normal day-to-day trips. However, this improvement in sample size is at the cost of poorer control over the collection method. For example, while the same device may be used, the speed of travel may vary greatly, and while travelling at high speed in a vehicle not all Wi-Fi Access Points present may be collected. However, recent quantitative evaluation has assessed this impact, finding that opportunistic wardriving collects at least 60% of the total number of Wi-Fi Access Points in a specific area [47]. Thus, this demonstrates there are still benefits to this approach as this disadvantage is not prohibitive to potentially collecting representative data.

*National statistics* are released on wireless networks collected via survey methods providing insight on current adoption and usage of digital technologies by citizens and businesses. Importantly, these statistics are often disaggregated based on the factors which affect adoption and usage. For example, population statistics are often broken down based on the respondents age, socio-economic status, region and settlement type (urban or rural). Equally, for businesses these statistics are usually reported by industrial sector and size (e.g. number of employees). However, the drawback to these statistics is that they are generally nationally representative, while not being provided at the local level. Additionally, the underlying survey data may not be available for access to protect the privacy of respondents. However, such data could still be used in deductive modeling approaches to provide local and national insight into the total number of Wi-Fi APs, particularly in terms of providing an estimated upper bound.

Having evaluated potential data sources for wireless networks, there is unfortunately no fully comprehensive truth dataset on Wi-Fi assets. Despite this, we do have several different data sources and methods which enable estimates of Wi-Fi assets to be derived. Indeed, a deductive modeling approach using national statistics does allow estimates to be obtained for Wi-Fi density by generalizing survey data. This then allows us to evaluate sampling methods, such as opportunistic wardriving.

## 3.  Method

We propose a method which consists of three main steps. Firstly, Step 1 involves using opportunistic wardriving to develop a self-collected geolocated Wi-Fi AP dataset. Next, Step 2 consists of developing a predictive model for Wi-Fi density using national statistics. Finally, Step 3 comparatively evaluates the different quantitative datasets on Wi-Fi APs from both Step 1 and 2, in order to identify statistical considerations with the wardriving method.

### 3.1. Collecting and processing wardriving data

An opportunistic wardriving sample is collected using the WiGLE app on an Android Google Pixel 4. Although the use of smartphones to collect AP information could lead to a loss of location accuracy due to less sophisticated antennas, this is mitigated by ensuring a level of data aggregation which reduces this effect [51], as discussed in detail later in this method. The smartphone approach is also representative of how most users now collect this data. Once complete, the data are processed based on the sequence of steps illustrated in *Figure 1*. This begins by exporting the collected data as .kml files and processing the results using the open-source codebase developed and released to accompany this paper entitled Wardriving Analytics for Wi-Fi with Python (wapy).

*Figure 1 Processing of self-collected wardriving Wi-Fi data*

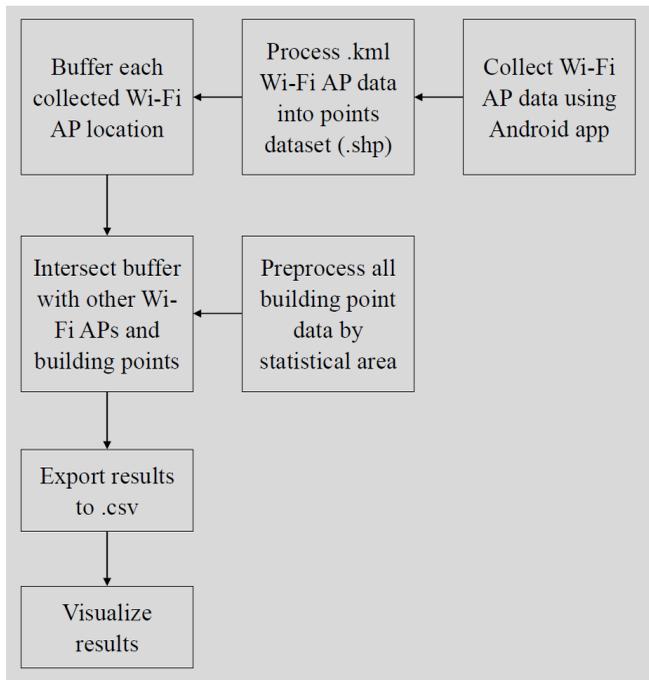

To obtain an understanding of Wi-Fi AP density it is necessary to count the number of assets in a specific geographic area. However, this processing presents statistical problems introduced by imposing artificial statistical boundaries. Within the spatial statistical literature, this is a well-documented problem known as the Modifiable Areal Unit Problem [52]–[54], whereby varying the position and size of statistical boundaries can have a significant impact on the yielded results. Often wardriving studies merely pick a single data aggregation scale without considering this issue [47]. Therefore, for each geolocated unique Wi-Fi network ID a circular buffer is added for a radius of 100-, 200- and 300-meters to explore the sensitivity of the data bounding box size. These sizes are justified given that AP geolocation detection using wardriving has been estimated to have a mean error of 32 meters [55], hence the values used here are large enough to keep this uncertainty to a minimum. Within each circular buffer, the intersecting number of premises points are also recorded, allowing premises count and density metrics to be estimated. Finally, the results are exported and analyzed for presentation in the results section.

### 3.2. Developing a predictive model for Wi-Fi Access Points

To predictively model the number of Wi-Fi APs we focus on the two main groups which adopt this technology to provide wireless Internet connectivity, households and businesses. As each group is quite different and thus has heterogeneous requirements for wireless connectivity, they must be treated separately.

Firstly, households usually purchase a fixed broadband subscription for use with a single wireless router. Although consumer Wi-Fi mesh networks are available for households, generally few have them, with most homes having a single Wi-Fi AP. Evidence supports the hypothesis that household density is a strong predictor of Wi-Fi AP density [56].

Secondly, businesses use different types of connectivity depending on their size. For example, small businesses frequently use consumer broadband products, much like households, and most probably only have a single Wi-Fi AP. This generally provides adequate service for those businesses with under ten employees. However, for those larger businesses with more than ten employees (or managed office spaces), they are much more likely to use a commercial Ethernet product providing symmetrical bandwidth (e.g. via fiber). The number of Wi-Fi APs each business has will then be positively correlated with the floor area of the premises. A single Wi-Fi AP usually serves between 20-200 square meters, so a business with offices spanning 1000 square meters could have between 5-20 Wi-Fi APs. Additionally, not all the floor area of a business may have Wi-Fi APs deployed, with this uncertainty resulting in an unknown model parameter.

With these factors in mind, a predictive modeling method is proposed in *Figure 2* using available national statistical data for Britain. The approach is based on the adoption estimates released for households by Ofcom from the Technology Tracker [57] and for businesses by the UK Office for National Statistics [58]. Both are based on survey methods and are designed to be statistically representative of the relative populations surveyed. The method produces disaggregated estimates of adoption which when aggregated match national statistics.

Population data for statistical output areas (N=8463) (*areas*) are taken [59] and used to allocate 'geotypes' which reflect different settlement patterns for urban (>7959 inhabitants km²), suburban (>782 inhabitants km²) and rural (<782 inhabitants km²) locations. The urban-rural definition is based on the segmentation used by the telecommunication regulator, in the Long Run Incremental Cost model used for supporting regulatory decisions relating to wireless communications [60]. This method is consistent with other studies in the literature [56].

The household input data are developed via a microsimulation approach which estimates the demographic composition of households in each local statistical area. Such data estimation methods are common in the demographics literature with the resultant data being able to inform demand for infrastructure services such as telecommunications or energy. For further information many papers in the literature provide a detailed explanation of the development of these demand forecasts [61]–[64].

*Figure 2 Predictive modeling method*

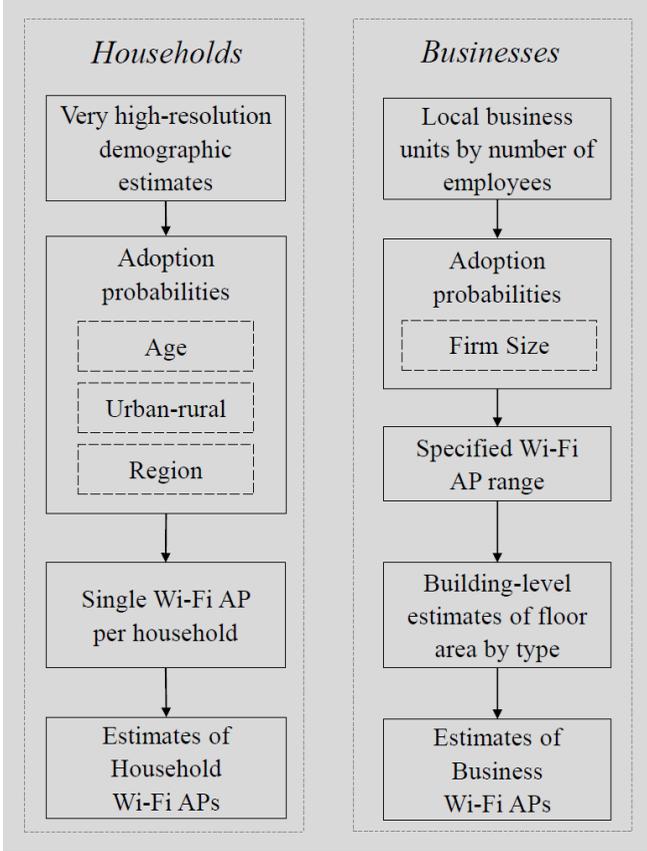

Let $i$ be the index for each individual resident within a household, allowing the level of broadband adoption ($B$) to be estimated first, and then within those the level of Wi-Fi adoption ($W$) can subsequently be estimated. The head of the household is treated as the oldest member and this is used to derive the age adoption probability ($A_i^B$) for the $i$th individual for fixed broadband adoption ($B$), along with the relevant probabilities for the region ($R_i^B$) and settlement pattern ($S_i^B$). The estimated mean adoption value ($\hat{p}_i^B$) for the $i$th individual is then used to estimate the take-up of fixed broadband, as illustrated in equation (1).

$$\hat{p}_i^B = \frac{A_i^B + R_i^B + S_i^B}{3} \tag{1}$$

Consequently, out of those individuals who have adopted fixed broadband, the level of Wi-Fi adoption is estimated ($W$), based on equation (2). Again, the head of the household is treated as the oldest member and this is used to derive adoption probabilities for age ($A_i^W$), region ($R_i^W$) and settlement pattern ($S_i^W$), enabling estimation of the mean adoption value ($p_i^W$).

$$\widehat{p_i^W} = \frac{A_i^W + R_i^W + S_i^W}{3} \tag{2}$$

Let $r$ be a random number between 0 and 1 taken from a uniform distribution, and using the algorithm in Figure 3 it is possible to estimate the number of residential Wi-Fi APs in an area.

*Figure 3 Wi-Fi adoption algorithm pseudocode*

```
1.  For every area in areas:
2.      For every individual i in area:
3.      ---------------------------------------------
4.          Assign a probability using proxy variable r
5.      ---------------------------------------------
6.          Draw r from uni. dist. in range(0,1):
7.              if 0 < r < p̂_i^B:
8.                  Set fixed broadband adoption a^B = 1
9.                  Draw r from uni. dist. in range(0,1):
10.                 if 0 < r < p̄_i^W:
11.                     Set a^W = 1
12.                 Else:
13.                     Set a^W = 0
14.             Else:
15.                 Set a^B = 0
```

The business count data are then obtained from the ONS Nomis website which provides local area economic data [65]. This data states the number of businesses in each area based on the number of employees (micro, small, medium and large businesses). The approach ensures that the broadband adoption probability ($\hat{p}_i^B$) for the $i$th company with $n$ employees leads to national estimates which are equal to the ONS national estimate of adoption.

In the absence of statistical estimates on Wi-Fi adoption, we use the estimated floor area for all non-residential buildings to propagate a distribution through the developed model given an expectation about the Wi-Fi density, thus treating this problem as an exercise in uncertainty propagation. The aim is to capture a range given a set of extreme values. The estimated floor area in each local statistical unit is derived using the building footprint and number of floors. Next, the approximate number of employees in each size category is obtained (micro: 5, small: 25, medium: 150, large: 350, very large: 750) and used to derive an estimate of the floor area for each business category. Finally, the floor area of all businesses with broadband is obtained to inform the floor area available for Wi-Fi to potentially cover (using the number of employees as a weight to disaggregate the total floor area). Finally, the number of Wi-Fi APs is estimated given the adopted floor area value, divided by the average coverage area of each AP (with the AP coverage area being the unknown parameter needing exploration). We test 100-, 200- and 300-meter variants of this parameter, referred to in the codebase as low, baseline and high coverage area scenarios.

## 4. Evaluating Wi-Fi data results

Between February and August 2020, over half a million collected samples were obtained leading to 332,196 unique geospatially located Wi-Fi network IDs, mainly in Cambridge and London but also across the North and West of England. This provides a balance of data for urban, suburban and rural settlements, as visually illustrated in *Figure 4*. The data are typical of opportunistically gathered wardriving information, as is common in the literature.

*Figure 4 Geolocated unique Wi-Fi network IDs*

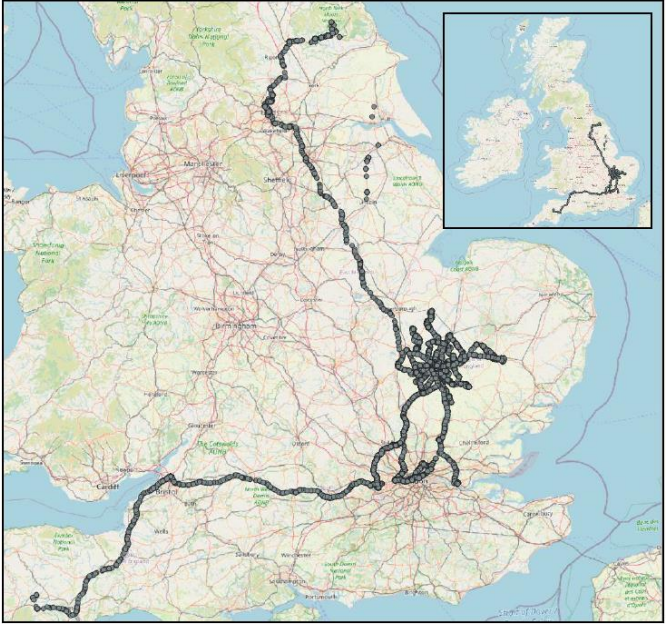

The density of Wi-Fi APs is reported in this section for both the wardriving and predictive modeling approaches. There are two main dimensions identified affecting the modeling of these assets which are therefore selected to present the results. This includes the urban-rural settlement type, and the buffer size measured around each Wi-Fi AP point, used to sample the number of proximate APs and premises in order to quantify density.

As visualized in *Figure 5*, the density of the Wi-Fi APs detected for urban areas via wardriving was highest when using the 100-meter buffer (0.03 km²) where approximately 3000 Wi-Fi APs km² were detected on average across the deciles, with this number raising to over 4000 km² in the densest areas. This contrasts with a flatter distribution of AP density counts when using the 200-meter (0.13 km²) and 300-meter (0.28 km²) buffers, with mean density values ranging approximately from 500-1500 Wi-Fi APs km².

For urban and suburban areas, the Wi-Fi AP density shows an interesting effect in that the number collected using wardriving is similar in both geotypes. This is despite urban areas being assumed *a priori* to have higher Wi-Fi AP density due to the increased density of businesses and households (with the predictive model estimates supporting this proposition). Such difference may be explained due to potential selection bias exhibited by wardriving urban environments leading to a loss of Wi-Fi APs located far above street level, and because propagation is often better in suburban areas due to less environmental clutter. Radio wave propagation in urban environments is usually poor due to large, dense buildings made of cement, steel and hardened glass. In contrast, suburban environments usually have lower building densities made with smaller quantities of construction materials (brick or wood) which are generally easier to propagate through.

Regarding the use of a 100-meter boundary (0.13 km²) in both urban and suburban areas, *Figure 5* suggests this buffer is too small leading to density inflation bias. Thus, Wi-Fi APs are being detected from outside of the area of analysis. This can be determined because the wardriving density is actually higher in many circumstances than the predictive model, which we treat as being closer to reality (as demonstrated next in the evaluation section 4.1). Hence, for the detection and density measurement of Wi-Fi APs, researchers should be favoring (at the very least) the baseline coverage area of 200 m to account for uncertainty introduced by wireless propagation.

In contrast, the wardriving results for lower population density settlements generally display much greater coherence with the predicted Wi-Fi AP densities. This is the case for the suburban and rural areas using the 200-meter (0.13 km²) and 300-meter (0.28 km²) buffers, with mean density values ranging approximately from 100-1000 Wi-Fi APs km². Having compared the results of the wardriving exercise with predictive values, the following section will use real building data to evaluate the predictive model.

### 4.1. Wi-Fi results validation

Data were combined from the University of Cambridge Wi-Fi register and the University of Cambridge Estates Management database in order to evaluate the predicted values against real Wi-Fi count data. The total number of Wi-Fi APs in operation per building was extracted and supplemented with the building floor area in square meters, allowing the Wi-Fi count per building to be compared with the baseline Wi-Fi AP coverage area (200m). Thus, in *Figure 5* buildings are ranked based on the total number of Wi-Fi APs present, and then a point-to-point plot is used to demonstrate the difference between the predicted and real values. *Figure 5* demonstrates that a general relationship exists. While there are deviations above or below the real values, buildings with smaller floor areas generally have fewer Wi-Fi APs and vice versa. This provides confidence in the model used, supporting the results produced.

## 5. Identification of statistical considerations

In this section we reflect on the results and identify important statistical considerations which are pertinent to future wardriving activities. As wardriving is a sampling-based method which aims to infer characteristics about a general population of Wi-Fi assets in a city or country, the method should adhere to gold-standard statistical design principles. Many existing studies in the literature fall short of this best practice, raising questions about the level of statistical rigor and robustness of the conclusions drawn. We highlight these issues here to inform the design of future wardriving studies.

*Figure 5 Wi-Fi APs: Predicted versus Wardriving*

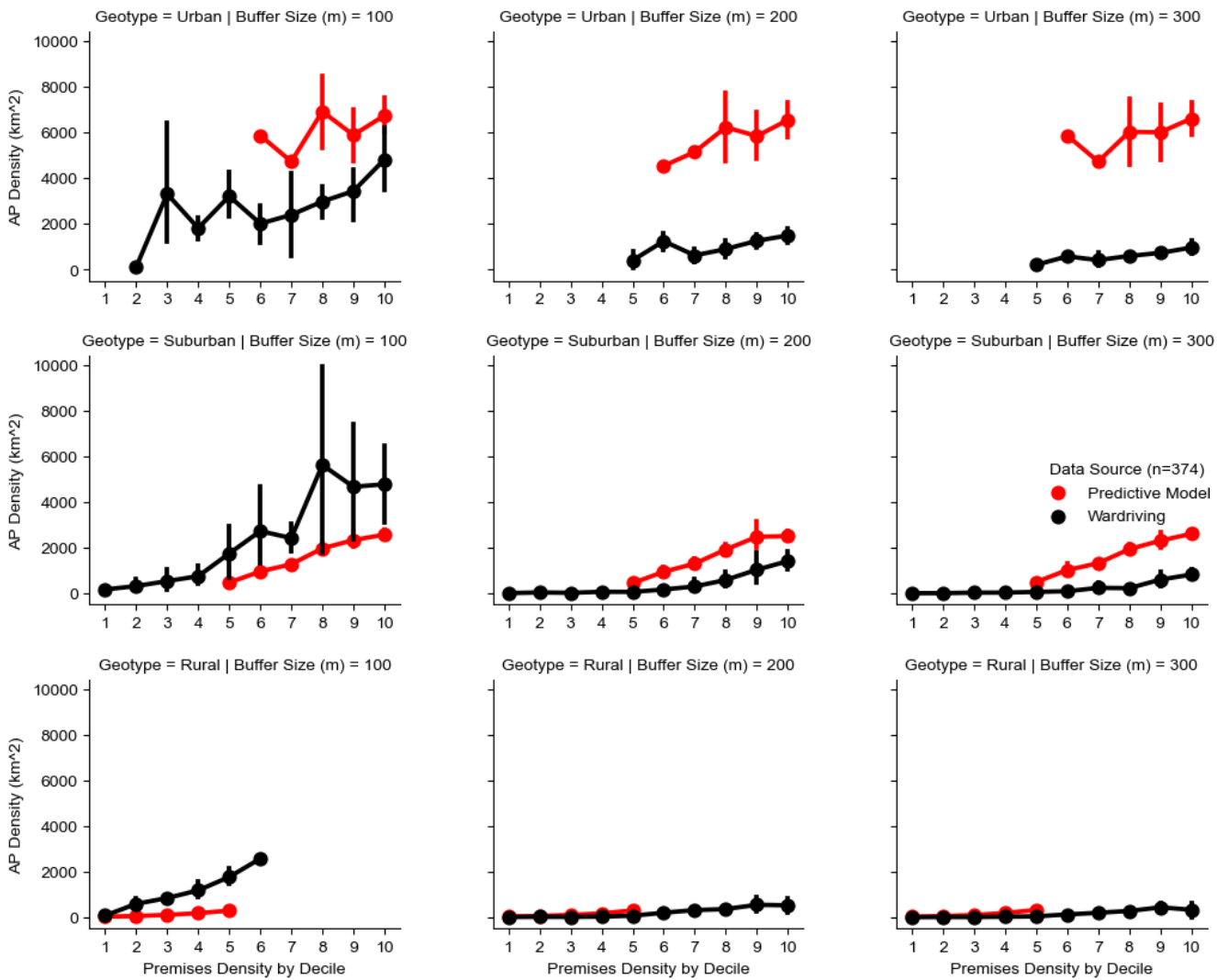

**Wardriving experiences selection bias** particularly in dense urban environments because the analyst is generally sampling along a very limited 3D plane, only sensing Wi-Fi APs on the first couple of floors of proximate buildings. This results from road vehicles being used to help survey larger geographic areas and therefore more Wi-Fi assets. Thus, it can be very challenging to sample highly populated urban areas where buildings are taller than a couple of floors and Wi-Fi AP signals cannot be easily sensed from the ground level (even at 2.4 GHz where propagation is generally better than 5 GHz). Researchers need to recognize that wardriving may be more suitable for suburban and rural environments, therefore additional design measures may need to be implemented to avoid select bias in dense urban environments. This could either include specific sampling of high-density locations using crowdsourced Wi-Fi data, such as via the WiGLE API, or considering the use of drone swarms to help better survey Wi-Fi APs in Three-Dimensional (3D) Space, when assets are far above street level ('wardroning').

**Wardriving exhibits a trade-off between rigor and scale** with subsequent statistical ramifications. The term 'expensive' is often used widely to reflect the ease of different sampling methods representing the amount of time and money required to obtain adequate sample sizes. While apps such as WiGLE may be free to download and use, the time expense required to survey a large area is considerable. Thus, there is a major trade-off in wardriving between dedicated surveying of small areas (which are expensive but attempt to control for local sampling bias), compared to opportunistic surveying (which provides greater spatial representation, but introduces statistical defects including the selection bias described above). Researchers need to select the correct type of wardriving sampling technique that best suits their research question, while being open about the shortcomings of different approaches.

**Wardriving studies often have issues with sample representativeness** which is an important consideration when interpreting the results of wardriving exercises. To enable external generalization when developing inferential statistics, suitably designed sampling strategies need to capture a representative set of entities from the total population in order to avoid biased results. Indeed, within the literature review of this paper, many studies identified have carried out relatively limited wardriving sampling exercises but then often attempted to generalize the results from the sample to the entire population. Without thorough statistical design, such generalizations beyond the sample are at best risky and at worst erroneous, therefore future studies need to ensure that this fundamental requirement for statistical analysis is met. Researchers should explicitly define the population of study if the results are to be generalized, and the quantitative statistics which demonstrate the collected sample is representative of this population need to be reported.

*Figure 6 Evaluation of the Wi-Fi AP Predictive Model*

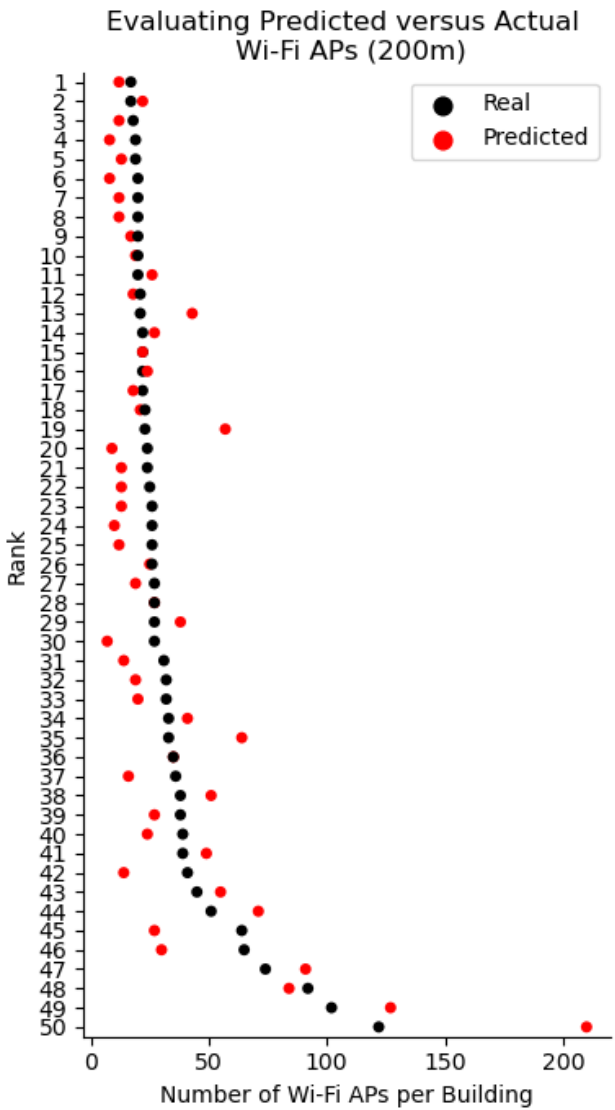

**Few wardriving assessments consider spatial statistical boundaries** which can have ramifications for the aggregation of wardriving data, given the fact this method is a spatial statistical process. Indeed, there are a variety of considerations which must be considered in order to avoid result inconsistencies, primarily related to the Modifiable Areal Unit Problem (MAUP). Any boundary specified around collected data is artificially imposed, whether using census boundary areas or 1 km² grid tiles. Therefore, two problems are potentially produced. Firstly, scale effects mean that the level of aggregation can have an influence on the Wi-Fi AP results. Using large geographic areas potentially smooth away spatially heterogenous data characteristics as they relate from the sample to the population. This contrasts with using high-resolution boundaries (e.g. 1 km² grid tiles) which retain much of this variance, leading to smaller losses in data fidelity. Secondly, the zoning of boundaries can produce uncertainty in the resulting data values, even when the underlying data are the same. For example, the gerrymandering of political boundaries is one particularly purposeful use of this statistical issue, but they could occur in wardriving studies unintentionally. To overcome these issues researchers should aim to retain as much data fidelity as possible when aggregating, as such an approach maintains the spatial heterogeneity in the collected data, preserving analytical insights useful for results interpretation.

**Wardriving is often poor at capturing temporal changes** driven by the fact that surveying can be expensive. Many of the wardriving assessments in the literature are spatial cross-sections of data collected via one-off opportunistic sampling exercises (the present paper included). The composition of Wi-Fi APs does not necessarily shift rapidly on a monthly basis, but over several years dynamic change could lead to differences in the number of APs and the type of Wi-Fi technology each asset is utilizing. In theory, the WiGLE API does allow access to data collected by other users but only a very limited number of queries are possible making it difficult to understand the temporal change in the distribution of Wi-Fi characteristics. Having identified and discussed these aspects of the wardriving method, conclusions will now be given in Section 6.

## 6. Conclusions

In this paper we demonstrate various statistical issues associated with the wardriving method. The approach was based on collecting a wardriving sample which passively sensed Wi-Fi APs, and then comparing these results to a predictive model developed using national statistics. The resulting estimates were evaluated for urban, suburban and rural areas, as well as for different sized buffer radii (100, 200 or 300-meters).

In the discussion of the results we identify five important statistical issues which can affect wardriving data and make recommendations for how researchers should aim to control for these issues in the design of future studies. These issues include selection bias, the trade-off between statistical rigor and scale, sample representativeness, the Modifiable Areal Unit Problem, and the challenges of capturing temporal information. By taking account of these factors we hope that future wardriving exercises will be able to provide more rigorous and robust statistical assessments of Wi-Fi APs. The codebase developed to reach these conclusions has been released open source and is available from the Wardriving Analytics for Wi-Fi with Python (wawpy) online repository.

## Acknowledgment

The authors wish to thank Facebook Connectivity Lab for an open science research grant on Wi-Fi sensing. We also thank Alexander Cox and Leigh Graves for helping to provide data to evaluate the predictive model.

## References

[1] H. A. Omar, K. Abboud, N. Cheng, K. R. Malekshan, A. T. Gamage, and W. Zhuang, 'A Survey on High Efficiency Wireless Local Area Networks: Next Generation WiFi', *IEEE Communications Surveys Tutorials*, vol. 18, no. 4, pp. 2315–2344, Fourthquarter 2016, doi: 10.1109/COMST.2016.2554098.

[2] E. J. Oughton, W. Lehr, K. Katsaros, I. Selinis, D. Bubley, and J. Kusuma, 'Revisiting Wireless Internet Connectivity: 5G vs Wi-Fi 6', *arXiv:2010.11601 [cs]*, Oct. 2020, Accessed: Dec. 28, 2020. [Online]. Available: http://arxiv.org/abs/2010.11601.

[3] G. Naik, J.-M. Park, J. Ashdown, and W. Lehr, 'Next Generation Wi-Fi and 5G NR-U in the 6 GHz Bands: Opportunities and Challenges', *IEEE Access*, vol. 8, pp. 153027–153056, 2020, doi: 10.1109/ACCESS.2020.3016036.

[4] J. D. Manton *et al.*, 'Development of an open technology sensor suite for assisted living: a student-led research project', *Interface Focus*, vol. 6, no. 4, p. 20160018, Aug. 2016, doi: 10.1098/rsfs.2016.0018.

[5] J. Eriksson, H. Balakrishnan, and S. Madden, 'Cabernet: vehicular content delivery using WiFi', in *Proceedings of the 14th ACM international conference on Mobile computing and networking*, San Francisco, California, USA, Sep. 2008, pp. 199–210, doi: 10.1145/1409944.1409968.

[6] R. Jedermann, T. Pötsch, and C. Lloyd, 'Communication techniques and challenges for wireless food quality monitoring', *Philosophical Transactions of the Royal Society A: Mathematical, Physical and Engineering Sciences*, vol. 372, no. 2017, p. 20130304, Jun. 2014, doi: 10.1098/rsta.2013.0304.

[7] J. P. Lynch, 'An overview of wireless structural health monitoring for civil structures', *Philosophical Transactions of the Royal Society A: Mathematical, Physical and Engineering Sciences*, vol. 365, no. 1851, pp. 345–372, Feb. 2007, doi: 10.1098/rsta.2006.1932.

[8] M. Mack, P. Dittmer, M. Veigt, M. Kus, U. Nehmiz, and J. Kreyenschmidt, 'Quality tracing in meat supply chains', *Philosophical Transactions of the Royal Society A: Mathematical, Physical and Engineering Sciences*, vol. 372, no. 2017, p. 20130308, Jun. 2014, doi: 10.1098/rsta.2013.0308.

[9] K. Soga and J. Schooling, 'Infrastructure sensing', *Interface Focus*, vol. 6, no. 4, p. 20160023, Aug. 2016, doi: 10.1098/rsfs.2016.0023.

[10] D. Wu, Q. Liu, Y. Li, J. A. McCann, A. C. Regan, and N. Venkatasubramanian, 'Adaptive Lookup of Open WiFi Using Crowdsensing', *IEEE/ACM Trans. Networking*, vol. 24, no. 6, pp. 3634–3647, Dec. 2016, doi: 10.1109/TNET.2016.2533399.

[11] R. Zhang *et al.*, 'A New Environmental Monitoring System Based on WiFi Technology', *Procedia CIRP*, vol. 83, pp. 394–397, Jan. 2019, doi: 10.1016/j.procir.2019.04.088.

[12] Z. Zou, Q. Chen, I. Uysal, and L. Zheng, 'Radio frequency identification enabled wireless sensing for intelligent food logistics', *Philosophical Transactions of the Royal Society A: Mathematical, Physical and Engineering Sciences*, vol. 372, no. 2017, p. 20130313, Jun. 2014, doi: 10.1098/rsta.2013.0313.

[13] Z. Fu, J. Xu, Z. Zhu, A. X. Liu, and X. Sun, 'Writing in the Air with WiFi Signals for Virtual Reality Devices', *IEEE Transactions on Mobile Computing*, vol. 18, no. 2, pp. 473–484, Feb. 2019, doi: 10.1109/TMC.2018.2831709.

[14] T. Kerdoncuff, A. Blanc, and N. Montavont, 'Using Empirically Validated Simulations to Control 802.11 Access Point Density', in *2017 IEEE Wireless Communications and Networking Conference (WCNC)*, Mar. 2017, pp. 1–6, doi: 10.1109/WCNC.2017.7925888.

[15] K. Kawamura, A. Inoki, S. Nakayama, K. Wakao, and Y. Takatori, 'Cooperative control of 802.11ax access parameters in high density wireless LAN systems', in *2019 IEEE Wireless Communications and Networking Conference (WCNC)*, Apr. 2019, pp. 1–6, doi: 10.1109/WCNC.2019.8885425.

[16] S. Stefanatos and A. Alexiou, 'Access Point Density and Bandwidth Partitioning in Ultra Dense Wireless Networks', *IEEE Transactions on Communications*, vol. 62, no. 9, pp. 3376–3384, Sep. 2014, doi: 10.1109/TCOMM.2014.2351820.

[17] S. Kirkpatrick, 'Once the Internet can measure itself', *Philosophical Transactions of the Royal Society A: Mathematical, Physical and Engineering Sciences*, vol. 374, no. 2062, p. 20140437, Mar. 2016, doi: 10.1098/rsta.2014.0437.

[18] A. W. T. Tsui, W.-C. Lin, W.-J. Chen, P. Huang, and H.-H. Chu, 'Accuracy Performance Analysis between War Driving and War Walking in Metropolitan Wi-Fi Localization', *IEEE Transactions on Mobile Computing*, vol. 9, no. 11, pp. 1551–1562, Nov. 2010, doi: 10.1109/TMC.2010.121.

[19] H. Berghel, 'Wireless infidelity I: war driving', *Commun. ACM*, vol. 47, no. 9, pp. 21–26, Sep. 2004, doi: 10.1145/1015864.1015879.

[20] Y.-L. Kim, 'Seoul's Wi-Fi hotspots: Wi-Fi Access Points as an indicator of urban vitality', *Computers, Environment and Urban Systems*, vol. 72, pp. 13–24, Nov. 2018, doi: 10.1016/j.compenvurbsys.2018.06.004.

[21] M. Seufert, C. Moldovan, V. Burger, and T. Hoßfeld, 'Applicability and limitations of a simple WiFi hotspot model for cities', in *2017 13th International Conference on Network and Service Management (CNSM)*, Nov. 2017, pp. 1–7, doi: 10.23919/CNSM.2017.8255985.

[22] F. Lyu, L. Fang, G. Xue, H. Xue, and M. Li, 'Large-

Scale Full WiFi Coverage: Deployment and Management Strategy Based on User Spatio-Temporal Association Analytics', *IEEE Internet of Things Journal*, vol. 6, no. 6, pp. 9386–9398, Dec. 2019, doi: 10.1109/JIOT.2019.2933266.

[23] Y. Tian, B. Huang, B. Jia, and L. Zhao, 'Optimizing WiFi AP Placement for Both Localization and Coverage', in *Algorithms and Architectures for Parallel Processing*, Cham, 2018, pp. 489–503, doi: 10.1007/978-3-030-05057-3_37.

[24] X. Du, K. Yang, and D. Zhou, 'MapSense: Mitigating Inconsistent WiFi Signals Using Signal Patterns and Pathway Map for Indoor Positioning', *IEEE Internet of Things Journal*, vol. 5, no. 6, pp. 4652–4662, Dec. 2018, doi: 10.1109/JIOT.2018.2797061.

[25] M. Maity, B. Raman, and M. Vutukuru, 'TCP Download Performance in Dense WiFi Scenarios: Analysis and Solution', *IEEE Transactions on Mobile Computing*, vol. 16, no. 1, pp. 213–227, Jan. 2017, doi: 10.1109/TMC.2016.2540632.

[26] R. Muhendra, A. Rinaldi, M. Budiman, and Khairurrijal, 'Development of WiFi Mesh Infrastructure for Internet of Things Applications', *Procedia Engineering*, vol. 170, pp. 332–337, Jan. 2017, doi: 10.1016/j.proeng.2017.03.045.

[27] L. Shillington and D. Tong, 'Maximizing wireless mesh network coverage', *International Regional Science Review*, vol. 34, no. 4, pp. 419–437, 2011.

[28] V. Q. Son and N. T. A. Khoa, 'Evaluation of Full-Mesh Networks for Smart Home Applications', in *2019 International Symposium on Electrical and Electronics Engineering (ISEE)*, Oct. 2019, pp. 73–78, doi: 10.1109/ISEE2.2019.8920920.

[29] M. Stellin, S. Sabino, and A. Grilo, 'LoRaWAN Networking in Mobile Scenarios Using a WiFi Mesh of UAV Gateways', *Electronics*, vol. 9, no. 4, Art. no. 4, Apr. 2020, doi: 10.3390/electronics9040630.

[30] J. Shen, J. Cao, and X. Liu, 'BaG: Behavior-aware Group Detection in Crowded Urban Spaces using WiFi Probes', *IEEE Transactions on Mobile Computing*, pp. 1–1, 2020, doi: 10.1109/TMC.2020.2999491.

[31] K. Gyöngyösi, P. J. Varga, and Z. Illési, 'WLAN heat mapping in hybrid network', in *2017 IEEE 14th International Scientific Conference on Informatics*, Nov. 2017, pp. 94–97, doi: 10.1109/INFORMATICS.2017.8327228.

[32] WiGLE, 'WiGLE: Wireless Network Mapping', 2020. https://wigle.net/ (accessed Jan. 10, 2020).

[33] A. Dagelić, M. Čagalj, T. Perković, and M. Biloš, 'Towards linking social media profiles with user's WiFi preferred network list', *Ad Hoc Networks*, vol. 107, p. 102244, Oct. 2020, doi: 10.1016/j.adhoc.2020.102244.

[34] Z. Illési, A. Halász, and P. J. Varga, 'Wireless Networks and Critical Information Infrastructure', in *2018 IEEE 12th International Symposium on Applied Computational Intelligence and Informatics (SACI)*, May 2018, pp. 000255–000260, doi: 10.1109/SACI.2018.8441023.

[35] R. C. Shah and C. Sandvig, 'SOFTWARE DEFAULTS AS DE FACTO REGULATION The case of the wireless internet', *Information, Communication & Society*, vol. 11, no. 1, pp. 25–46, Feb. 2008, doi: 10.1080/13691180701858836.

[36] A. Valenzano, D. Mana, C. Borean, and A. Servetti, 'Mapping WiFi measurements on OpenStreetMap data for Wireless Street Coverage Analysis', in *Free and Open Source Software for Geospatial (FOSS4G) Conference Proceedings*, 2016, vol. 16, no. 1, p. 5.

[37] R. Banakh and A. Piskozub, 'Attackers' Wi-Fi Devices Metadata Interception for their Location Identification', in *2018 IEEE 4th International Symposium on Wireless Systems within the International Conferences on Intelligent Data Acquisition and Advanced Computing Systems (IDAACS-SWS)*, Sep. 2018, pp. 112–116, doi: 10.1109/IDAACS-SWS.2018.8525538.

[38] S. Gupta, B. S. Chaudhari, and B. Chakrabarty, 'Vulnerable analysis using war driving and security intelligence', in *2016 International Conference on Inventive Computation Technologies (ICICT)*, Aug. 2016, vol. 3, pp. 1–5, doi: 10.1109/INVENTIVE.2016.7830165.

[39] P. R. Lutui, O. Tete'imoana, and G. Maeakafa, 'An analysis of personal wireless network security in Tonga: A study of Nuku'alofa', in *2017 27th International Telecommunication Networks and Applications Conference (ITNAC)*, Nov. 2017, pp. 1–4, doi: 10.1109/ATNAC.2017.8215409.

[40] H. Valchanov, J. Edikyan, and V. Aleksieva, 'A Study of Wi-Fi Security in City Environment', *IOP Conf. Ser.: Mater. Sci. Eng.*, vol. 618, p. 012031, Oct. 2019, doi: 10.1088/1757-899X/618/1/012031.

[41] S. Konomi, T. Sasao, S. Hosio, and K. Sezaki, 'Exploring the Use of Ambient WiFi Signals to Find Vacant Houses', in *Ambient Intelligence*, Cham, 2017, pp. 130–135, doi: 10.1007/978-3-319-56997-0_10.

[42] Z. Akram, M. A. Saeed, and M. Daud, 'Wardriving and its Application in Combating Terrorism', in *2018 1st International Conference on Computer Applications Information Security (ICCAIS)*, Apr. 2018, pp. 1–5, doi: 10.1109/CAIS.2018.8442035.

[43] C. L. Leca, 'Overview of Romania 802.11 wireless networks security', in *2017 9th International Conference on Electronics, Computers and Artificial Intelligence (ECAI)*, Jun. 2017, pp. 1–4, doi: 10.1109/ECAI.2017.8166386.

[44] M. Maráczi, 'Wardriving in Eger', in *2019 IEEE 13th International Symposium on Applied Computational Intelligence and Informatics (SACI)*, May 2019, pp. 000127–000130, doi: 10.1109/SACI46893.2019.9111489.

[45] E. Nasr, M. Jalloul, J. Bachalaany, and R. Maalouly, 'Wi-Fi Network Vulnerability Analysis and Risk Assessment in Lebanon', *MATEC Web Conf.*, vol. 281, p. 05002, 2019, doi: 10.1051/matecconf/201928105002.

[46] A. Sebbar, SE. Boulahya, G. Mezzour, and M. Boulmalf, 'An empirical study of WIFI security and

performance in Morocco - wardriving in Rabat', in *2016 International Conference on Electrical and Information Technologies (ICEIT)*, May 2016, pp. 362–367, doi: 10.1109/EITech.2016.7519621.

[47] A. Bandong, C. Felizardo, C. A. M. Festin, and W. M. Tan, 'Opportunistic Wardriving Through Neighborhood Public Utility Vehicles as an Alternative to Crowdsourcing and Dedicated Wardriving for Wireless Network Data Collection', in *2018 IEEE International Conference on Internet of Things and Intelligence System (IOTAIS)*, Nov. 2018, pp. 66–72, doi: 10.1109/IOTAIS.2018.8600890.

[48] F. J. Díaz, M. Robles, P. Venosa, N. Macia, and G. Vodopivec, 'Wardriving: an experience in the city of La Plata', 2008.

[49] D. Dobrilovic, Z. Stojanov, S. Jäger, and Z. Rajnai, 'A method for comparing and analyzing wireless security situations in two capital cities', *Acta Polytechnica Hungarica*, vol. 13, no. 6, pp. 67–86, 2016.

[50] A. Sarrafzadeh and H. Sathu, 'Wireless LAN security status changes in Auckland CBD: A case study', in *2015 IEEE International Conference on Computational Intelligence and Computing Research (ICCIC)*, Dec. 2015, pp. 1–6, doi: 10.1109/ICCIC.2015.7435676.

[51] A. Achtzehn, L. Simić, M. Petrova, and P. Mähönen, 'IEEE 802.11 Wi-Fi Access Point density estimation with capture-recapture models', in *2015 International Conference on Computing, Networking and Communications (ICNC)*, Feb. 2015, pp. 153–159, doi: 10.1109/IC-CNC.2015.7069333.

[52] A. S. Fotheringham and D. W. S. Wong, 'The Modifiable Areal Unit Problem in Multivariate Statistical Analysis', *Environ Plan A*, vol. 23, no. 7, pp. 1025–1044, Jul. 1991, doi: 10.1068/a231025.

[53] S. Openshaw, 'The modifiable areal unit problem', *Quantitative geography: A British view*, pp. 60–69, 1981.

[54] D. W. S. Wong, 'The Modifiable Areal Unit Problem (MAUP)', in *WorldMinds: Geographical Perspectives on 100 Problems: Commemorating the 100th Anniversary of the Association of American Geographers 1904–2004*, D. G. Janelle, B. Warf, and K. Hansen, Eds. Dordrecht: Springer Netherlands, 2004, pp. 571–575.

[55] M. Kim, J. J. Fielding, and D. Kotz, 'Risks of Using AP Locations Discovered Through War Driving', in *Pervasive Computing*, Berlin, Heidelberg, 2006, pp. 67–82, doi: 10.1007/11748625_5.

[56] A. Achtzehn, L. Simić, P. Gronerth, and P. Mähönen, 'Survey of IEEE 802.11 Wi-Fi deployments for deriving the spatial structure of opportunistic networks', in *2013 IEEE 24th Annual International Symposium on Personal, Indoor, and Mobile Radio Communications (PIMRC)*, Sep. 2013, pp. 2627–2632, doi: 10.1109/PIMRC.2013.6666591.

[57] Ofcom, 'Ofcom Nations and Regions Technology Tracker - H1 2020', 2020. https://www.ofcom.org.uk/__data/assets/pdf_file/0037/194878/technology-tracker-2020-uk-data-tables.pdf (accessed Jun. 10, 2020).

[58] ONS, 'E-commerce and ICT activity, UK - Office for National Statistics', 2018. https://www.ons.gov.uk/businessindustryandtrade/itandinternetindustry/bulletins/ecommerceandictactivity/2018 (accessed Jul. 25, 2020).

[59] ONS, 'Open Geography Portal - Output Areas', 2020. https://geoportal.statistics.gov.uk/search?collection=Dataset (accessed Sep. 08, 2020).

[60] Ofcom, 'Mobile call termination market review 2015-18', Ofcom, London, 2015. Accessed: Oct. 27, 2016. [Online]. Available: https://www.ofcom.org.uk/consultations-and-statements/category-1/mobile-call-termination-14.

[61] J. W. Hall *et al.*, 'Strategic analysis of the future of national infrastructure', *Proceedings of the Institution of Civil Engineers - Civil Engineering*, pp. 1–9, Nov. 2016, doi: 10.1680/jcien.16.00018.

[62] E. J. Oughton and T. Russell, 'The importance of spatio-temporal infrastructure assessment: Evidence for 5G from the Oxford–Cambridge Arc', *Computers, Environment and Urban Systems*, vol. 83, p. 101515, Sep. 2020, doi: 10.1016/j.compenvurbsys.2020.101515.

[63] A. P. Smith, 'humanleague: a C++ microsynthesis package with R and python interfaces', *Journal of Open Source Software*, vol. 3, no. 25, p. 629, May 2018, doi: 10.21105/joss.00629.

[64] C. Thoung *et al.*, 'Future demand for infrastructure services', in *The future of national infrastructure: A system-of-systems approach*, Cambridge: Cambridge University Press, 2016.

[65] ONS, 'Nomis - Official Labour Market Statistics', 2020. https://www.nomisweb.co.uk/ (accessed Mar. 22, 2020).

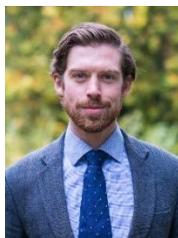

**EDWARD J. OUGHTON** received the M.Phil. and Ph.D. degrees from the Clare College, University of Cambridge, U.K., in 2010 and 2015, respectively. He later held research positions at both Cambridge and Oxford. He is currently an Assistant Professor with George Mason University, Fairfax, VA, USA, developing open-source research software to analyze digital infrastructure deployment strategies. He received the Pacific Telecommunication Council Young Scholars Award in 2019, Best Paper Award 2019 from the Society of Risk Analysis, and the TPRC Charles Benton Early Career Award 2021.

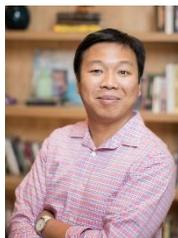

**JULIUS KUSUMA** received his B.S. from Purdue University in 1999, M.S. from the University of California at Berkeley in 2001, and Ph.D. from the Massachusetts Institute of Technology in 2006, all in Electrical Engineering and Computer Science. He was a visiting scientist at École Polytechnique Fédérale de Lausanne in 2001 and 2004. In 2006-2018 he was Principal Scientist at Schlumberger-Doll Research, working on subsurface and subsea connectivity technologies. He joined Facebook Connectivity in 2018 where he is working on technologies for rural and wireless connectivity. He has been awarded 20 US patents. He received the Ted Rappaport Scholarship in 1999, the Demetri Angelakos Memorial Award in 2001, and the MIT Presidential Fellowship in 2001.

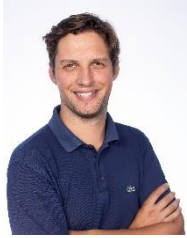

**THIBAULT PEYRONEL** is Thibault Peyronel received his MSc from Ecole Polytechnique in 2007 and his PhD in Physics from MIT in 2013. He was a Postdoctoral Fellow in Physics at Harvard University from 2013 to 2015. He joined Facebook Connectivity in 2015 as a Research Scientist, working on hardware solutions, data science and machine learning

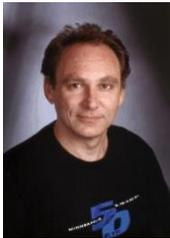

**JON CROWCROFT** received the degree in physics from Trinity College, University of Cambridge, in 1979, and the M.Sc. degree in computing and the Ph.D. degree from UCL, in 1981 and 1993, respectively. He has been a Marconi Professor of communications systems with the Computer Laboratory, University of Cambridge, since 2001. He was involved in the area of the Internet support for multimedia communications for more than 30 years. His three main topics of interest have been scalable multicast routing, practical approaches to traffic management, and the design of deployable end-to-end protocols. His current research interests include opportunistic communications, social networks, and techniques and algorithms to scale infrastructure-free mobile systems. He is a Fellow of the Royal Society, the ACM, the British Computer Society, the IET, and the Royal Academy of Engineering.